\documentclass{nature}
\usepackage[english]{babel}
\newcommand{\onlinecite}[1]{\hspace{-1 ex} \nocite{#1}\citenum{#1}} 
\usepackage{amsmath}
\usepackage{amsfonts}
\usepackage{amssymb}
\usepackage{graphicx}
\usepackage{amstext}
\begin{document}
\title{Universality of Non-equilibrium Fluctuations in Strongly Correlated Quantum Liquids}

%% Notice placement of commas and superscripts and use of &
%% in the author list

\author{  Meydi Ferrier$^{1,2}$, Tomonori Arakawa$^{1}$, Tokuro Hata$^{1}$, Ryo Fujiwara$^{1}$, Rapha{\"e}lle Delagrange$^{2}$, Rapha{\"e}l Weil$^{2}$, Richard Deblock$^{2}$, Rui Sakano$^{3}$, Akira Oguri$^{4}$, Kensuke Kobayashi$^{1}$ }

\maketitle

%\affiliation{$^{1}$ Osaka University, $^{2}$ LPS, Univ. Paris-Sud, CNRS, UMR 8502, F-91405 Orsay Cedex, France,$^{3}$ ISSP, $^{4}$ Osaka City University}

\begin{affiliations}
 \item  Department of Physics, Graduate School of Science, Osaka University, 1-1 Machikaneyama, Toyonaka, 560-0043 Osaka, Japan
 \item  Laboratoire de Physique des Solides, CNRS, Univ. Paris-Sud, Universit{\'e} Paris Saclay, 91405 Orsay Cedex, France
 \item  The Institute for Solid State Physics, The University of Tokyo, Kashiwa, Chiba 277-8581, Japan 
 \item  Department of Physics, Osaka City University, Sumiyoshi-ku, Osaka 558-8585, Japan
\end{affiliations}

\begin{abstract}
%For Nature, the abstract is really an introductory paragraph set
%in bold type.  This paragraph must be ``fully referenced'' and
%less than 180 words for Letters.  This is the thing that is
%supposed to be aimed at people from other disciplines and is
%arguably the most important part to getting your paper past the
%editors.  End this paragraph with a sentence like ``Here we
%show...'' or something similar.

Interacting quantum many-body systems constitute a fascinating playground for researchers since they form quantum liquids with correlated ground states and low-lying excitations, which exhibit universal behaviour\cite{MahanBook}. In fermionic systems, such quantum liquids are realized  in helium-3 liquid, heavy fermion systems\cite{MahanBook}, neutron stars and  cold gases\cite{Bloch2012}. Their properties in the linear-response regime have been successfully described by the theory of Fermi liquids\cite{MahanBook}.  However, non-equilibrium properties beyond this regime have still to be established and remain a key issue of many-body physics. Here, we show a precise experimental demonstration of Landau Fermi-liquid theory extended to the non-equilibrium regime in a 0-D system.  Combining transport and ultra-sensitive current noise measurements, we have unambiguously identified  the SU(2) \cite{VanderWiel2000} and SU(4) \cite{Jarillo-herrero2005,Makarovski2007,Cleuziou2013,Schmid2015,Delattre2009} symmetries of quantum liquid  in a carbon nanotube tuned in the universal Kondo regime. We find that, while the electronic transport is well described by the free quasi-particle picture around equilibrium \cite{Nozieres1974}, a two-particle scattering process due to residual interaction shows up in the non-equilibrium regime \cite{Sela2006,Oguri2005,Gogolin2006,Mora2008,Mora2009,Sakano2011,Zarchin2008,Delattre2009,Yamauchi2011}. By using the extended Fermi-liquid theory, we obtain the interaction parameter ``Wilson ratio''\cite{Wilson1975} $R=1.9 \pm 0.1$ for SU(2) and $R=1.35 \pm 0.1$ for SU(4) as well as the corresponding effective charges, characterizing the quantum liquid behaviour. This result, in perfect agreement with theory\cite{Sela2006,Gogolin2006,Mora2008,Mora2009,Sakano2011}, provides a strong quantitative experimental background for further developments of the many-body physics. Moreover, we discovered a new scaling law for the effective charge, signalling as-yet-unknown universality in the non-equilibrium regime. Our method to address quantum liquids through their non-equilibrium noise paves a new road to tackle the exotic nature of quantum liquids out-of-equilibrium in various physical systems \cite{Egger2009}.

\end{abstract}

%Then the body of the main text appears after the intro paragraph.
%Figure environments can be left in place in the document.
%\verb|\includegraphics| commands are ignored since Nature wants
%the figures sent as separate files and the captions are
%automatically moved to the end of the document (they are printed
%out with the \verb|\end{document}| command. However, tables must
%be manually moved to the end of the document, after the addendum.

The Kondo effect \cite{Kondo1964} is a typical example of quantum many body effect, where a localised spin is screened by the surrounding conduction electrons at low temperature to form a unique correlated ground state. The Kondo state is well described by the Fermi-liquid theory at equilibrium \cite{MahanBook,Nozieres1974}, which makes it an ideal test-bed to go beyond.  To unveil the universal behaviour of non-equilibrium Fermi-liquid \cite{Oguri2005}, we have used the current fluctuations or shot noise in a Kondo-correlated nanotube quantum dot\cite{Egger2009}.

\begin{figure}
  \centering
  \includegraphics[width=16cm]{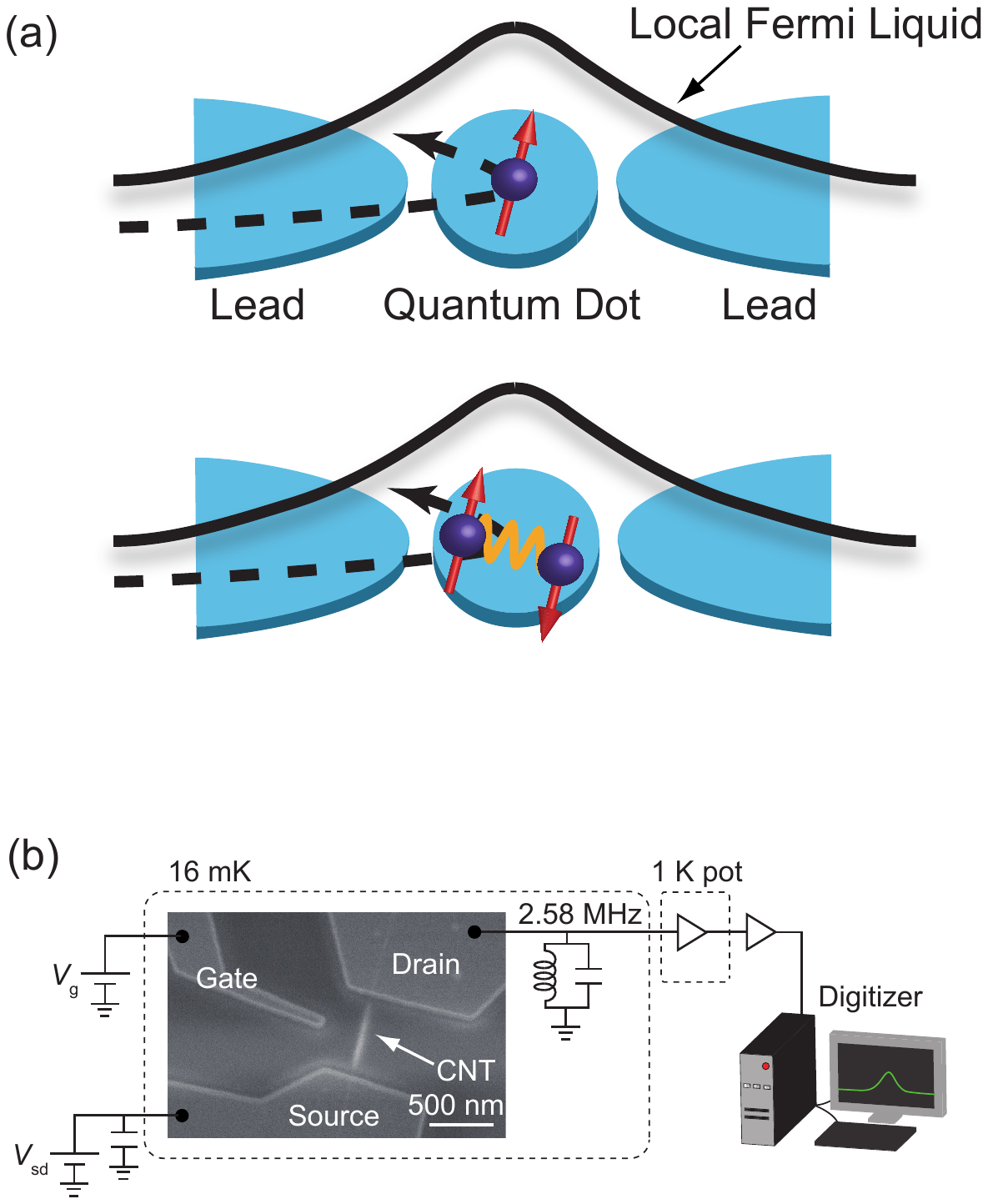}
 %\vspace {-0.5cm}
  \caption{
  \textbf{Illustration of the interaction effect in the quantum liquid and design of the sample.}
  \textbf{a)} Scheme of a quantum dot coupled to two leads. If a single spin is confined on the dot and the coupling with the leads is strong enough, the Kondo effect occurs.  The full black line represents the extension of the Kondo state where electrons form a strongly correlated Fermi-liquid. Our work consists in a collision experiment. We inject electrons through this state and observe via the current fluctuations which quasi-particles are backscattered. Upper part) At low energy, this liquid can be understood as an ensemble of non-interacting quasi-particles. During a collision, only single quasi-particles are excited and backscattered by an incoming electron. Lower part) At higher energy the incoming electron can excite two quasi-particles of the Fermi-liquid through the residual interaction between quasi-particles (orange meander line). The scattered current consists then of single or double quasi-particles shots which increase the current fluctuations. The effective charge $e^*$ is a measure of the probability for these two processes. The Wilson ratio $R$ quantifies the strength of the interaction between two quasi-particles.   \textbf{b)} Sample and experimental set-up. SEM image of a carbon nanotube connected to metallic leads on a silicon wafer. It is connected to the current and noise measurement set-up through a resonant circuit fixed on the mixing chamber of the dilution fridge. The gate voltage $V_g$ tunes the number of electrons ($N$) in the dot and the coupling between dot and leads.}
\label{fig1}
\end{figure}

When electrons are transmitted through this system, the scattering induces the shot noise, which sensitively reflects the nature of the quasi-particles\cite{Blanter2000}, as shown in the upper panel of Fig.\ref{fig1}1a. A remarkable prediction of the non-equilibrium Fermi-liquid theory is that the residual interaction creates an additional scattering of two particles which enhances the noise (see the lower panel of Fig.\ref{fig1}1a) \cite{Sela2006,Gogolin2006,Mora2008,Mora2009,Sakano2011}. This two-particle scattering is characterized by an effective charge $e^*$  larger than $e$ (electron charge). This value, closely related to the Wilson ratio, is universal for the Fermi-liquid in the Kondo regime since it only depends on the symmetry group of the system \cite{Mora2008,Sakano2011,Mora2009}. While some aspects of the Kondo-associated noise were reported \cite{Zarchin2008,Delattre2009,Yamauchi2011}, a rigorous, self-consistent treatment in a regime where universal results apply is at the core of the present work. Actually, by tuning a nanotube quantum dot from the spin degenerate SU(2) Kondo regime to the spin-orbit degenerate SU(4) one, the noise is proven to contain distinct signatures of these two symmetries, confirming theoretical developments of Fermi-liquid theory out of equilibrium.

\begin{figure}
\centering
\includegraphics[height=18cm]{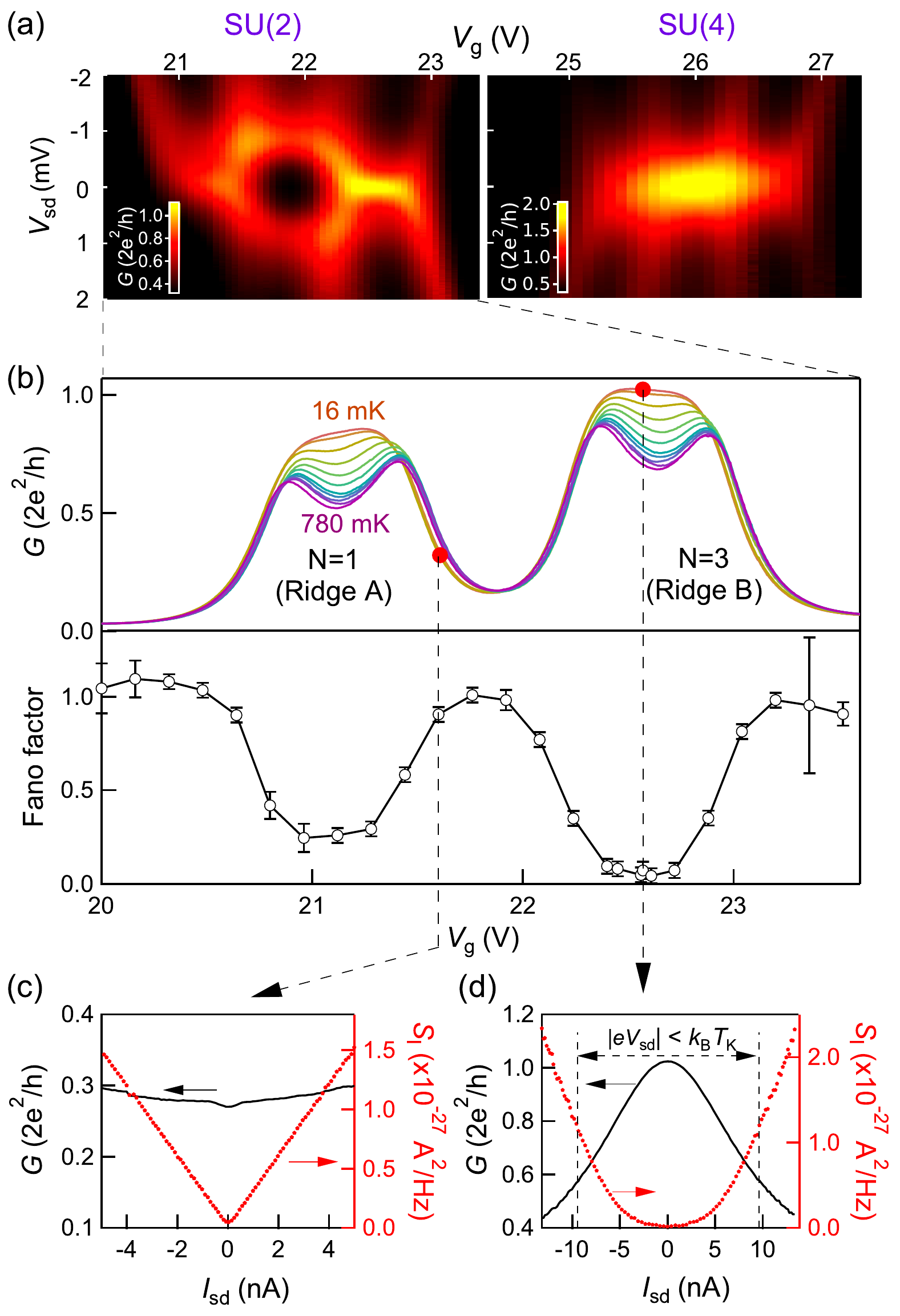}
\caption{

\textbf{Figure 2: Stability diagram and shot noise for the SU(2) Kondo effect at \boldmath$T=16\ $mK.} \textbf{a)} 2D plot of the conductance as a function of $V_{sd}$ and $V_g$. Kondo ridges appear as bright horizontal lines for $V_{sd}=0$. In the SU(2) region two ridges show up for $N=1$ and $N=3$ electrons. For SU(4) the Kondo ridge exists for $N=1,2\ $and $3$ electrons. Note that the conductance scale is $2e^2/h$ for SU(2) whereas it is $2\times 2e^2/h$ for SU(4). An in-plane magnetic field of $0.08\ $T is applied to suppress superconductivity in the contacts. \textbf{b) Upper panel:} Cross-sections of the conductance of the SU(2) Kondo region at zero bias as a function of $V_g$ for different temperatures. Two Kondo ridges A and B are clearly visible.  \textbf{Lower panel:} Fano factor extracted from the linear part of the current noise in the regime $eV_{sd} \ll k_BT_K$ ($I_{sd}\leq 5\ $nA) using the definition $S_i=2eF|I_{sd}|$. \textbf{c)} Conductance (black line) and noise (red dots) as a function of $I_{sd}$ for the blockaded region ($N=2$). The noise is linear with $|I_{sd}|$ with a slope around $2e$. Outside the Kondo ridge due to the very small transmission, the quantum dot is close to the tunneling regime characterized by a full poissonian noise with $F=1$. \textbf{d)} Conductance (black line) and noise (red dots) as a function of $I_{sd}$ on the Kondo ridge ($N=3$). The slope of $S_i$ around $I_{sd}=0$ is almost $0$ due to the perfect transmission ($F = 0.06$).}
\label{figCB}
\end{figure}

In our experiment, we measured the conductance and current noise through a carbon nanotube quantum dot grown by chemical vapour deposition\cite{Kasumov2007} on an oxidized undoped silicon wafer and connected with a Pd ($6\ $nm)/Al($70\ $nm) bilayer. The distance between the contacts is 400 nm and a side gate electrode is deposited to tune the potential of the quantum dot (see Fig\ref{fig1} 1b). To measure accurately the shot noise, our sample is connected to a resonant ($2.58\ $MHz) LC circuit. The signal across this resonator is measured with a home-made cryogenic low-noise amplifier\cite{Arakawa2013}. Figure \ref{figCB}2a presents the image-plot of the differential conductance  of the sample ($G$) at temperature $T=16\ $mK as a function of source-drain voltage ($V_{sd}$) and gate voltage ($V_g$).  This stability diagram shows the four-fold degenerated Coulomb diamonds specific to carbon nanotubes. The spectrum consists in successive four-electron shells. We note $N=0, 1, 2, 3\ $ the number of electrons in the last shell.  Remarkably, the SU(2) and SU(4) Kondo ridges\cite{Cleuziou2013} emerge as horizontal bright regions (high conductance) at $V_{sd}=0$. 

%\section{results}
%\subsection{Characterization}
For the moment, we concentrate on the SU(2) region.
 A cut of the conductance at $V_{sd}=0$ is represented in the upper panel of Fig.~\ref{figCB}2b. Two Kondo ridges appear as plateaux where $G$ is maximum for the filling $N=1\ $ (ridge A) and $N=3$ (ridge B), whereas $G$ decreases to almost zero for even $N$. In addition,  the ridge B is flat and the unitary limit is achieved: the conductance reaches the quantum of conductance $G_Q=2e^2/h$ which is a signature of perfect Kondo effect in a dot with symmetric coupling to the leads.
The Kondo temperature ($T_K$) is $1.6\pm 0.05\ $K in the centre of this ridge (see supplementary).

The current noise $S_i$ as well as $G$ are plotted on the Figs.~\ref{figCB}2c and \ref{figCB}2d for $N=2$ and $N=3$ as a function of the source-drain current $I_{sd}$. Outside the Kondo ridge ($N=2$) $S_i$ is linear with $|I_{sd}|$, whereas on the ridge the shot noise is  flat around $I_{sd}=0\ $ and  enhanced at high current when the energy of incoming electrons approaches a fraction of $T_K$. At high voltage ($eV_{sd}\gg k_BT_K$), a linear behaviour is recovered with $S_i=2e|I_{sd}|$. To analyse the low energy properties, we have extracted the Fano factor ($F$) which is defined as: $S_i=2eF|I_{sd}|$ from a linear fit at low current. $F$ varies approximately from zero on the Kondo ridge to $F=1$ outside as shown in the bottom panel of Fig.\ref{figCB}2b. Indeed at very low energy, as the free quasi-particle picture of the Fermi-liquid theory teaches us,  the conductance and the noise for a multichannel conductor can be written as a function of transmission $T_i$ for each channel $i=1, 2, 3\cdots$ \cite{Blanter2000}:
\begin{equation}
G=G_Q\sum T_i\ \quad\mathrm{and}\quad\ F=\frac{G_Q}{G}\sum T_i(1-T_i).
\label{Fano}
\end{equation}
 For the SU(2) symmetry, transport occurs through one single channel yielding $G=G_QT_1$ and $F=1-T_1$.   On the Kondo ridge, the conductance is $G=G_Q$ yielding $T_1=1$ and $F=1-T_1=0$. This is a direct signature of the Kondo resonance which allows a perfect transmission and thus no partition of quasi-particles near equilibrium. In the Coulomb blockade regime for even $N$, the transport is blocked ($T_1\ll 1$) yielding $F\approx 1$ indicating that transport occurs through tunnelling events resulting in a conventional Poissonian noise.
%\subsection{SU2 Kondo effect}

\begin{figure}
\centering
\includegraphics[width=15cm]{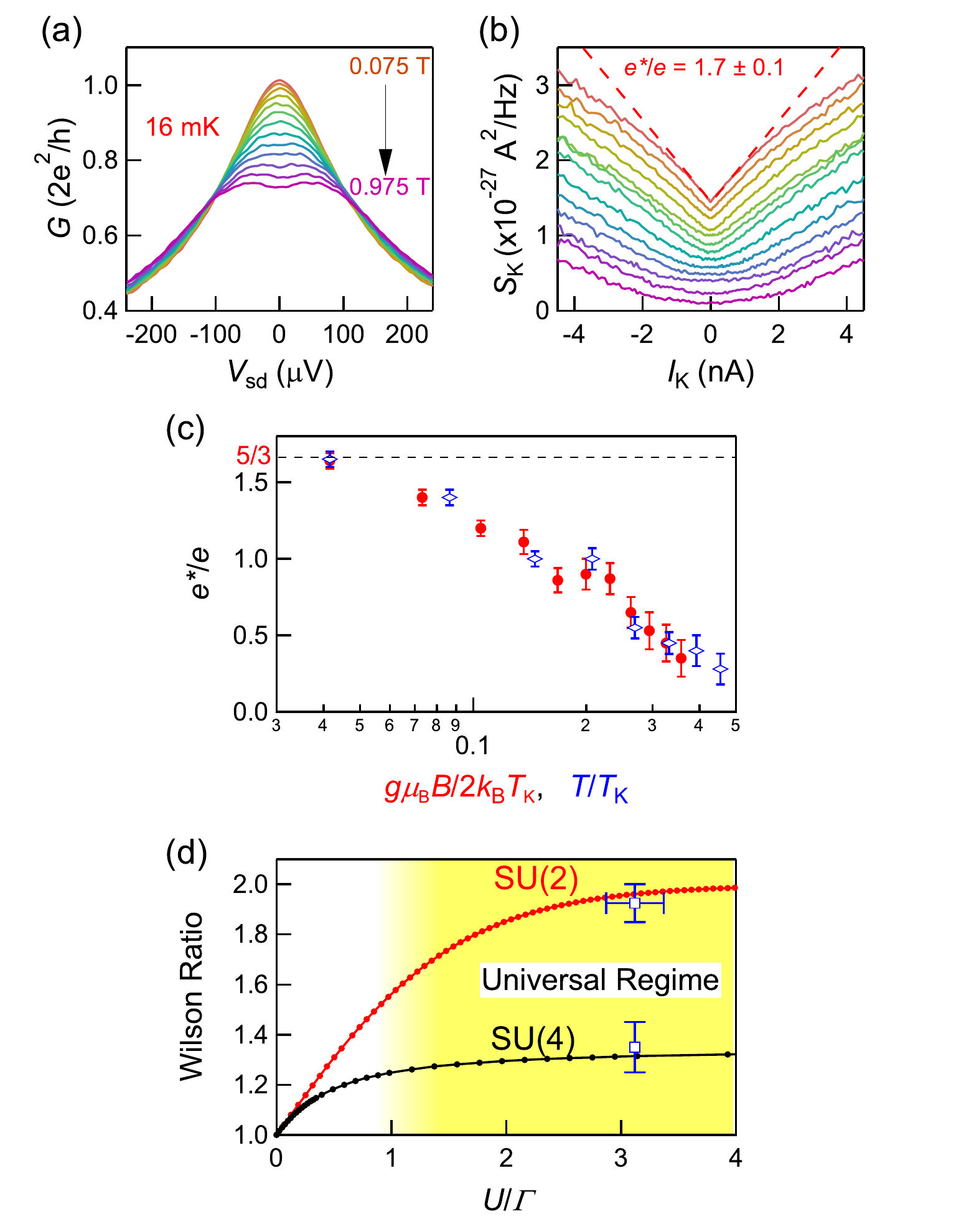}
\caption{

\textbf{Figure 3: Evolution of Kondo correlations with magnetic field and temperature.}
\textbf{a)} Evolution of the differential conductance from $0.075\ $T to $0.975\ $T with a step of $0.075\ $T. %The Kondo peak is splitted in two Zeeman satellites from which we can extract the g-factor of the nanotube g=2 (see supplementary).
\textbf{b)} Kondo excess current noise $S_K$ as a function of backscattering  current $I_K$ (see text for definition) for the same fields. The effective charge $e^*$ is extracted from the slope $\frac{S_K}{2eI_K}$ at low current. The red dotted-line shows result of the linear fit yielding $e^*/e=1.7 \pm 0.1$. The curves are shifted for clarity.
\textbf{c)} Effective charge as a function of $T/T_K$ and $\frac{g\mu_BB}{2k_BT_K}$ in a semi-log plot. Both  can be described by the same line which is a signature of a scaling law for $e^*(T,B)$ and suggest a logarithmic behaviour.
\textbf{d)} Wilson ratio as a function of $U/\Gamma$ calculated with the formula from [\onlinecite{Sakano2011}]. For SU(2), vertical error bars  represent the value for $R$ extracted from $e^*$ and from the scaling analysis of the conductance shown in the supplementary. Horizontal error bar represents the value for $U/\Gamma$ extracted independently from the fit of Kondo temperature shown in the supplementary. The crossing on the theoretical curve shows the agreement of our measurement with the extension of  Fermi-liquid theory out of equilibrium. For SU(4), the value of R is extracted from $e^*$. The yellow part of the graph represents the region of universality : all the properties depend on a single parameter $T_K$ for a given symmetry group \cite{Oguri2005}.}
\label{Bfield}
\end{figure}

At higher voltages ($0<eV_{sd}\leq k_BT_K/2$) we concentrate only on the non-linear terms for current and noise by subtracting the linear part. We  defined $S_K = S_i-2eF|I_{sd}| $ and the backscattered current $I_K=G(0)V_{sd}-I_{sd}$ \cite{Mora2008,Sela2006}. These quantities are related through the effective charge\cite{Sela2006,Gogolin2006,Mora2008,Sakano2011a} $e^*$: 
\begin{equation}
S_K=2e^* |I_K|.
\label{SK}
\end{equation} 
This effective charge does not imply an exotic charge as in the quantum Hall regime but it is related to the probability that one particle or two particles are backscattered in the Fermi-liquid.
Figures \ref{Bfield}3a and \ref{Bfield}3b show the evolution of the conductance on ridge B as a function of magnetic field $B$ and the corresponding noise $S_K$ as a function of $I_K$, respectively. The temperature dependence is analysed in details in the supplementary. The effective charge is directly given by the slope at  low current $I_K$ ($eV_{sd} < k_BT_K$) yielding for the lowest field and temperature $e^*/e=1.7\pm 0.1$. This result is in good agreement with theory \cite{Sela2006,Gogolin2006}, which predicts $e^*/e=5/3\approx 1.67$ corresponding to equal probabilities for one or two-particle scattering. %It is a direct measurement of the two particles scattering induced by the residual interaction between quasiparticles \cite{Sela2006,Mora2009}. 
%Also we clearly see that this effect is destroyed by increasing magnetic field  or temperature. 
Figure 3c represents the evolution of $e^*$ with magnetic field and temperature. On this graph $e^*$ is represented as a function of the reduced scales $\frac{T}{T_K}$ for temperature or $\frac{g\mu_BB}{2k_BT_K}$ for the field where $g=2$ is the Land{\'e} factor and $\mu_B$ the Bohr magneton (see supplementary). All the data points seem to fall on the same curve suggesting that $e^*$ obeys a logarithmic scaling law which has not yet been predicted.
 % as predicted in  [\onlinecite{Oguri2013}]. 

 The Wilson ratio $R$ has been extracted from the formula\cite{Gogolin2006,Sakano2011}:
\begin{equation}
\frac{e^*}{e}=\frac{1+9(n-1)\left(R-1\right)^2}{1+5(n-1)\left(R-1\right)^2}
\label{estar}
\end{equation}
 where $n$ characterizes the symmetry group SU($n$) of the Kondo state.
 This number $R$ is directly related to the ratio $U/\Gamma$  with $U$ the charging energy of the quantum dot and $\Gamma$ the coupling to the electrodes. It reflects the strength of the interaction in the Fermi-liquid and is the only parameter to characterize the system, going from $R=1$ in the non interacting case ($U=0$) to $R=2$ in the strong SU(2) Kondo limit ($U\rightarrow \infty$) (see Fig.\ref{Bfield}3d). Our value for $e^*$ yields $R= 1.95\pm 0.1$ ensuring strong interactions and thus universal regime.

%\subsection {scaling}
 For consistency, $R$ was independently extracted by fitting the evolution of conductance with $B, V_{sd}$ and $T$ at low excitation (see supplementary) without any assumption on $T_K$. This result, $R= 1.95 \pm 0.1$, perfectly agrees with the value extracted from $e^*$. Finally, from  the dependence of $T_K$ with gate voltage, we have independently extracted the values  $U=6\pm 0.5\ $meV  and $\Gamma=1.8\pm 0.2\ $meV. This consistency is illustrated in Fig.\ref{Bfield}3d where the two independent values for $R$ and $U/\Gamma$ cross on the theoretical curve.

The effect of asymmetric lead-dot coupling was tested on the ridge A where $G=0.85\ G_Q$. Asymmetry is defined by the factor $\delta$ such that $G(0)=(1-\delta)G_Q$. We have measured an effective charge $e^*/e=1.2 \pm 0.08$ in good agreement with Ref. [\onlinecite{Mora2009}] which predicts $\frac{e^*}{e}=\frac{5}{3}-\frac{8}{3}\delta=1.26$.

\begin{figure}%
    \centering
    \includegraphics[width=18cm]{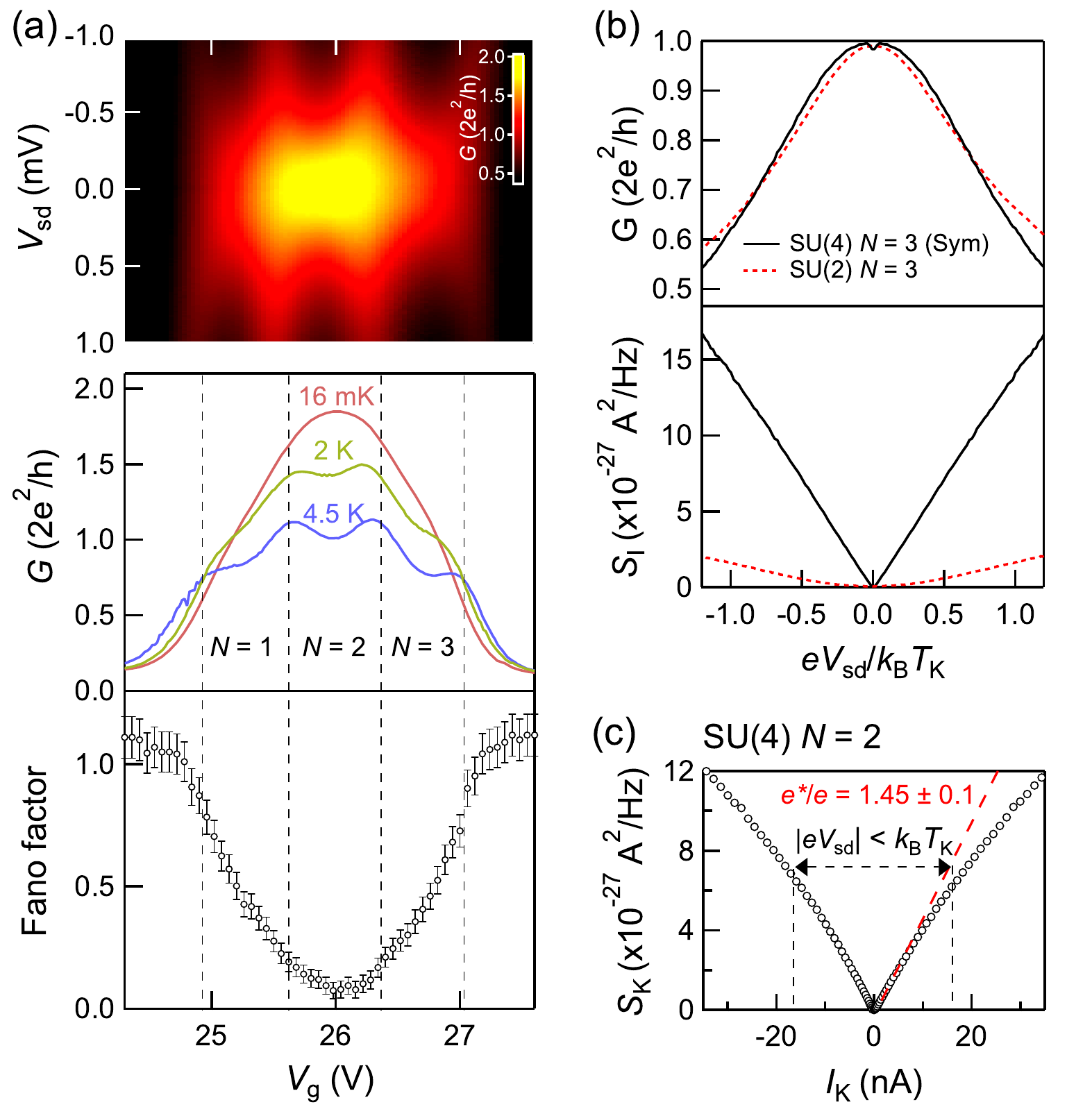}
    %\subfloat[label 1]{{\includegraphics[width=7cm]{fig4left} }}%
    %\qquad
    %\subfloat[label 2]{{\includegraphics[width=7.5cm]{fig4right} }}%
    \caption{

    \textbf{Figure 4: The SU(4) Kondo state and comparison with the SU(2) symmetry.} \textbf{a) top:} 2D plot of the conductance as a function of $V_{sd}$ and $V_g$ in the SU(4) state. \textbf{Middle}: cross-section of the conductance as a function of the gate voltage at $V_{sd}=0$ measured at different temperatures. $G$ reaches $1.8\times \frac{2e^2}{h}$ for the filling $N=2$. \textbf{Bottom:} Extracted Fano factor. As expected it varies from $F\approx 0$ for $N=2$ then $F=0.5$ for $N=1$ and finally $F=1$ in the tunneling regime (outside the Kondo ridge). \textbf{b)} Comparison of the conductance and the noise for SU(2) Kondo state and SU(4) at filling $N=3$. In both cases the conductance is the same and the two states are not distinguishable. At low voltage, the noise $S_i$ is qualitatively strongly different. For SU(2) the slope (Fano factor) is close to 0 since the transmission is almost perfect ($T_1=1$) whereas the linear part is very important in the SU(4) state since transport occurs through two half transmitting channels ($T_1=T_2=0.5$) yielding $F=0.5$. For clarity, since  asymmetry with respect to $V_{sd}$ appears in the SU(4) state at large voltage, we only compare the symmetrized part of the conductance and the noise for SU(4). We used the definition : $G^{sym}(V_{sd})=\frac{G(V_{sd})+G(-V_{sd})}{2}$ and $S_i^{sym}(V_{sd})=\frac{S_i(V_{sd})+S_i(-V_{sd})}{2}$.  \textbf{c)} Kondo excess noise for the SU(4) symmetry at filling $N=2$. A linear fit at low current using formula \ref{SK} gives $e^*/e=1.45\pm 0.1$ yielding $R=1.35\pm 0.1$.}
    \label{SU4}%
\end{figure}

Now, what can we learn from the SU(4) symmetry emerging in the right part of Fig.~\ref{figCB}2a? A zoom is plotted in the upper panel of Fig.\ref{SU4}4a and the cross-section is displayed in the middle part. Since spin and orbital degrees of freedom are degenerated, two channels contribute to transport and Kondo resonance emerges at every filling factors $N=1$, $2$ and $3$ electrons \cite{Jarillo-herrero2005,Makarovski2007,Cleuziou2013,Schmid2015}. At odd filling, the channels are half transmitted ($T_1=T_2=0.5$) yielding the same conductance $G_Q$ as in the SU(2) symmetry. However  for $N=2$, current is transmitted through two perfect channels ($T_1=T_2=1$) increasing the conductance up to $G=2\ G_Q$. In this region the conductance hardly depends on temperature up to $800$ mK reflecting a large $T_K$ as expected for the SU(4) symmetry\cite{Galpin2005}. This is confirmed by the full width of the curve $G(V_{sd})$ which gives $T_K \approx 11\ $K for $N=2$ and $T_K\approx 17\ $K for $N=1$ or $3$.

The first result to emphasize is that the linear part of the current noise is qualitatively different from SU(2) and is a powerful experimental tool to distinguish the two symmetries \cite{Mora2009,Egger2009}.
The upper part of Fig. \ref{SU4}4b represents the conductance at $N=3$ for SU(2) and SU(4) as a function of the rescaled voltage $eV_{sd}/k_BT_K$. The two curves are barely distinguishable. However the current noise displayed on the lower part of Fig. \ref{SU4}4b is qualitatively different since it is almost zero for the SU(2) symmetry whereas it is linear with $|I_{sd}|$ for SU(4). The linear noise is  one order of magnitude stronger for SU(4) than for SU(2). Indeed, the Fano factor for two channels such that $T_1=T_2=0.5$ is $F=0.5$ for the SU(4) symmetry whereas for SU(2) a single channel with $T_1=1$ yields $F=0$ . The complete evolution of $F$ is summarized in the lower part of Fig. \ref{SU4}4a. It changes from $F\approx 1$ outside the Kondo ridge to $F=0.5$ for $N=1$ and $3$ and reaches $F=0.07$ for $N=2$. This confirms that in the $N=2$ SU(4) case, transport takes place through two almost perfect channels ($T_1=T_2=1$) without partition  yielding $F\approx 0$.

 Finally we discuss $e^*$ and $R$ for the SU(4) symmetry at half filling ($N=2$). $S_K$ and $I_K$ have been computed by using the same procedure as explained for SU(2). The result, displayed on Fig. \ref{SU4}4c, gives  $e^*/e=1.45\pm 0.1$ and $R=1.35 \pm 0.1$ from Eqn~(\ref{estar}). The decreasing value for $R$ and $e^*$ reflects the increasing number of degenerated states. When degeneracy increases, electrons correlations become weaker and the non-interacting value is recovered  for SU(n) when $n\rightarrow \infty$. The values for $e^*$ and $R$ are in good agreement with Refs. [\onlinecite{Mora2009,Sakano2011}] (also see Fig.~3d) confirming that  non-equilibrium Fermi-liquid theory can be extended to more exotic classes of Fermi-liquids. %The extraction of the Kondo scattering outside the electron-hole symmetric  point ($N=1$ and $3$) needs much more data treatment and theoretical analysis which will be done in further works.
 
  Our experimental results emphasize three important points for the theory of Fermi-liquids. First, in the linear regime they can be described as free quasi-particles as the spirit of the theory teaches us. This allows to clearly  distinguish the different symmetry class of Fermi-liquid through shot noise measurements, while it is hardly possible from the conductance. Second, by probing the nonlinear noise, we have shown that out-of-equilibrium the residual interaction between quasi-particles shows up and creates a peculiar two-particle scattering, which bares the signature of the correlated nature of quantum liquids. Finally the newly discovered out-of-equilibrium scaling law should trigger theoretical developments to deepen our understanding of universal behaviours in these liquids.

%\begin{figure}
%\caption{Each figure legend should begin with a brief title for
%the whole figure and continue with a short description of each
%panel and the symbols used. For contributions with methods
%sections, legends should not contain any details of methods, or
%exceed 100 words (fewer than 500 words in total for the whole
%paper). In contributions without methods sections, legends should
%be fewer than 300 words (800 words or fewer in total for the whole
%paper).}
%\end{figure}

\begin{methods}
%Put methods in here.  If you are going to subsection it, use
%\verb|\subsection| commands.  Methods section should be less than
%800 words and if it is less than 200 words, it can be incorporated
%into the main text.

%\subsection{Method subsection.}

%Here is a description of a specific method used.  Note that the
%subsection heading ends with a full stop (period) and that the
%command is \verb|\subsection{}| not \verb|\subsection*{}|.

The nanotube has been grown by a very low pressure CVD technique \cite{Kasumov2007}. We have deposited Fe catalyst by electron beam lithography on a non-doped silicon substrate and exposed it to $10\ $mbar of acetylene  during $9\ $s at a temperature of $900\ ^{\circ}$C. Nanotubes were located by SEM and contacts were designed by standard e-beam lithography. The metallic bilayer ($6\ $nm Pd/$70\ $nm Al) is deposited on the contacts by e-gun evaporation. An in-plane magnetic field $B=0.08\ $T is applied to suppress the superconductivity of the contacts. 

The shot noise is measured with a resonant set-up \cite{Arakawa2013}. The sample is connected to a LC circuit with a resonance frequency of $2.58\ $MHz thermalized on the mixing chamber of the dilution fridge. The power spectral density of the noise is obtained through the following procedure: amplifying the noise signal with a home-made amplifier fixed on the 1K pot of the dilution fridge and again at room-temperature, taking the time-domain signal by a digitizer (National Instruments PCI-5922) and performing the fast Fourier transformation of the data. The current noise of the sample is then extracted from the fit of the shape of the resonance in the frequency domain.

\end{methods}

%% Put the bibliography here, most people will use BiBTeX in
%% which case the environment below should be replaced with
%% the \bibliography{} command.

%\begin{thebibliography}{1}
%\bibitem{dummy} Articles are restricted to 50 references, Letters
%to 30.
% \bibitem{dummyb} No compound references -- only one source per
%reference.
%\end{thebibliography}
\bibliography{BibNat}
%\bibliography{library}
%% Here is the endmatter stuff: Supplementary Info, etc.
%% Use \item's to separate, default label is "Acknowledgements"

\begin{addendum}
\item[Correspondence] Correspondence and requests for materials
should be addressed to M. Ferrier.~(email: meydi.ferrier@u-psud.fr) or K. Kobayashi.~(email: kensuke@meso.phys.sci.osaka-u.ac.jp)
 \item We appreciate discussions with H. Bouchiat and R. Yoshii. This work was partially supported by a Grant-in-Aid for Scientific
 Research (S) (No. 26220711), JSPS KAKENHI (No. 26400319,  25800174 and 15K17680), Invitation Fellowships for Research in
 Japan from JSPS, Grant-in-Aid for Scientific Research on Innovative
 Areas "Fluctuation $\&$ Structure" (No. 25103003) and  "Topological Materials Science" (KAKENHI Grant No. 15H05854), the Program for
 Promoting the Enhancement of Research Universities from MEXT, and
 Yazaki Memorial Foundation for Science and Technology, the French program ANR DYMESYS (ANR2011-IS04-001-01) and ANR MASH (ANR-12-BS04-0016). KK acknowledges the stimulated discussion
 in the meeting of the Cooperative Research Project of RIEC, Tohoku
 University.
% \item[Competing Interests] The authors declare that they have no
%competing financial interests.
% \item[Correspondence] Correspondence and requests for materials
%should be addressed to A.B.C.~(email: myaddress@nowhere.edu).
\end{addendum}

%%
%% TABLES
%%
%% If there are any tables, put them here.
%%

%
%\begin{figure}
%\centering
%\includegraphics[width=\linewidth]{fig4.pdf}
%\caption{
%Stability diagram in the SU(4) Kondo state.}
%\label{SU4}
%\end{figure}

\end{document}